\documentclass{aa}

\newcommand\dosingle[1]{#1}  \newcommand\dodouble[1]{ } 
\newcommand\postrefereechanges[1]{#1}

\newcommand\nice[1]{#1}    \newcommand\subm[1]{}   
\subm{ \renewcommand\dosingle[1]{ }   \renewcommand\dodouble[1]{#1} }

\dodouble{ \documentclass[referee]{aa} }
\usepackage{epsf1990}  

\usepackage{url} 

\newcommand\zzz[2]{#2} 

\newcommand\SSS{Sect.~}

\nice{ \newcommand\mycaptionfont{\protect\footnotesize} }

\newcommand\kms{\,km\,s$^{-1}$}
\newcommand\centreline{\centerline}
\newcommand\gtapprox{\,\lower.6ex\hbox{$\buildrel >\over \sim$} \, }
\newcommand\ltapprox{\,\lower.6ex\hbox{$\buildrel <\over \sim$} \, }
\newcommand\propapprox{\,\lower.6ex\hbox{$\buildrel \propto\over \sim$} \, }

\newcommand\e{ {\scriptstyle \times} 10^}

\newcommand\arcs{\ifmmode {'' }\else $'' $\fi}     
\newcommand\arcm{\ifmmode {' }\else $' $\fi}       
\newcommand\ddeg{\ifmmode^\circ\else$^\circ$\fi}    

\newcommand\frtoday{Le\space\number\day\space\ifcase\month\or
  janvier\or f\'evrier\or mars\or avril\or mai\or juin\or
  juillet\or ao\^ut\or septembre\or octobre\or novembre\or 
d\'ecembre\fi\space \number\year}

\newcommand\hGpc{\mbox{$h^{-1}$ Gpc}}
\newcommand\hMpc{\mbox{$h^{-1}$ Mpc}}

\newcommand\rinj{{r}_{\mbox{\rm \small inj}}}  
\newcommand\rSLS{r_{\mbox{\rm \small SLS}}}  
\newcommand\Smax{S_{\mbox{\rm \small max}}}

\newcommand\Omm{\Omega_{\mbox{\rm \small m}}}
\newcommand\Omtot{\Omega_{\mbox{\rm \small tot}}}

\newcommand\lII{l^{\mbox{\sc II}}}
\newcommand\bII{b^{\mbox{\sc II}}}

\title{A Hint of Poincar\'e Dodecahedral Topology 
in the WMAP First Year Sky Map}

\author{Boudewijn F. Roukema\inst{1},
Bartosz Lew\inst{1},
Magdalena Cechowska\inst{1},
Andrzej Marecki\inst{1},
Stanislaw Bajtlik\inst{2}
}

\institute{Toru\'n Centre for Astronomy, N. Copernicus University,
ul. Gagarina 11, PL-87-100 Toru\'n, Poland 
\and
Nicolas Copernicus Astronomy Centre, 
ul. Bartycka 18, PL-00-716 Warsaw, Poland
}

\date{\frtoday}

\titlerunning{Dodecahedral Topology in WMAP}
\authorrunning{Roukema et al.}

\begin{document}

\abstract{
It has recently been suggested 
by \protect\nocite{LumNat03}{Luminet} {et~al.} (2003) that the WMAP data are better
matched by a geometry in which the topology is that of 
a Poincar\'e dodecahedral model and the curvature is ``slightly'' spherical,
rather than by an (effectively) infinite flat model. A general back-to-back
matched circles analysis by \protect\nocite{CSSK03}{Cornish} {et~al.} (2004) 
for angular radii in the range $25-90\ddeg$, using a correlation statistic
for signal detection, failed to support this. In this paper, a matched circles
analysis specifically designed to detect dodecahedral patterns of matched
circles is performed over angular radii in the range $1-40\ddeg$ 
on the one-year WMAP data. Signal detection is attempted via a 
correlation statistic and an rms difference
statistic.
Extreme value distributions of these statistics are calculated for 
one orientation of the 36$\ddeg$ `screw motion' (Clifford translation)
when matching circles, 
for the opposite screw motion, and for a zero (unphysical) rotation. 
The most correlated circles appear for circle 
radii of $\alpha =11\pm1 \ddeg$,
for the left-handed screw motion, but not for the right-handed one, nor 
for the zero rotation. 
The favoured six dodecahedral face centres 
in galactic coordinates are $(\lII,\bII)$ 
$\approx (252\ddeg,+65\ddeg), (51\ddeg,+51\ddeg),$
$(144\ddeg,+38\ddeg), (207\ddeg,+10\ddeg),$
$(271\ddeg,+3\ddeg), (332\ddeg,+25\ddeg)$ and their opposites. 
The six pairs of circles {\em independently} each
favour a circle angular radius of $11\pm1\ddeg$. 
The temperature fluctuations along the matched circles are plotted
and are clearly highly correlated.
Whether or not these six circle pairs centred on dodecahedral faces
match via a $36\ddeg$ rotation only due to unexpected statistical 
properties of the WMAP ILC map, or whether they match due to global
geometry, it is clear that the WMAP ILC map has some unusual statistical
properties which mimic a potentially interesting cosmological signal.
\keywords{cosmology: observations -- cosmic microwave background}
}

\maketitle

\dodouble{ \clearpage } 


\newcommand\tdodec{
\begin{table}
\caption{Sky positions of the six face centres for the dodecahedron
which shows excess values of the correlation statistic $S$. 
A face number $i$, galactic
longitude, latitude and estimated circle radius $\alpha$ (all in degrees)
are listed. The other 6 faces are directly opposite.
The orientation of the $36\ddeg$ screw motion between faces is left-handed.
\label{t-dodec}}
$$\begin{array}{c c c c} \hline 
\rule{0ex}{2.5ex}
i &
\lII \mbox{ in $\ddeg$} & \bII \mbox{ in $\ddeg$} 
& \alpha \mbox{ in $\ddeg$}\\ \hline 
   1 &     252.4 &  64.7    & 9.8 \\
   2 &      50.6  & 50.8   & 10.7 \\
   3 &     143.8  & 37.8 & 10.7 \\
   4 &     207.5  & 9.5 &   10.7 \\
   5 &     271.0  & 2.7 &  11.8 \\
   6 &     332.8   & 25.0 &    10.7 \\
\hline
\end{array}$$
\end{table}
}  

\newcommand\tbestshort{
\begin{table*}
\caption{
Axis positions of a 3-manifold candidate found close to a
long axis position listed in 
Table~\protect\ref{t-bestlong}, 
where the long axis ($Z_{T^2}$) 
is larger than the horizon diameter $2 R_H$ and 
$2\rinj \equiv 2R_H/10$ is the length of the two
short axes ($X_{T^2}$, $Y_{T^2}$). 
The KS probability of finding the observed temperature
differences given the 3-manifold as a null hypothesis is 
$P_{\mbox{\rm all}}$ or $P_{\mbox{\rm subs}}$, depending on whether
the full set of circles (where $N=150$ independent pairs are
assumed) or an evenly spaced
subset of $N=138$ pairs of circles, respectively, is used.
An ISW/systematic noise contribution of $x^2=0.3$ 
[eq.~(\protect\ref{e-xdefn})] is adopted. Statistics $\sigma,$ $d$ and
$S$ are defined in eqs~(\protect\ref{e-sigma}), 
(\protect\ref{e-dmean}) and (\protect\ref{e-corr}) respectively.  
\label{t-bestshort}}
$$\begin{array}{c c c c c c c c ccc} \hline 
 \multicolumn{2}{c}{\mbox{\rm long ($Z_{T_2}$)}} & 
\multicolumn{2}{c}{\mbox{\rm short ($X_{T_2}$)}} & 
\multicolumn{2}{c}{\mbox{\rm short ($Y_{T_2}$)}} & 
P(\mbox{\rm all}) & P(\mbox{\rm subs}) & 
\sigma & d & S \\
\mbox{$l^{{\sc II}}$} & \mbox{$b^{{\sc II}}$} & 
\mbox{$l^{{\sc II}}$} & \mbox{$b^{{\sc II}}$} & 
\mbox{$l^{{\sc II}}$} & \mbox{$b^{{\sc II}}$} \\ 
\hline 
280 & 37.5 & 184 & 8 & 264 & -51 & 0.40 & 0.19
& 1.61 & -0.006 & 0.21 \\
\hline
\end{array}$$
\end{table*}
}  

\newcommand\tworst{
\begin{table*}
\caption{Axis positions, statistics and null hypothesis 
probabilities,  
as for Table~\ref{t-bestshort},
for a $T^2$, $2\rinj=2R_H/10$ 
model strongly rejected using the
identified circles principle, in spite of the presence of 
a strong ISW effect, with $x^2=0.6$. 
\label{t-worst}}
$$\begin{array}{c c c c c c c c ccc} \hline 
 \multicolumn{2}{c}{\mbox{\rm long ($Z_{T_2}$)}} & 
\multicolumn{2}{c}{\mbox{\rm short ($X_{T_2}$)}} & 
\multicolumn{2}{c}{\mbox{\rm short ($Y_{T_2}$)}} & 
P(\mbox{\rm all}) & P(\mbox{\rm subs}) & 
\sigma & d & S \\
\mbox{$l^{{\sc II}}$} & \mbox{$b^{{\sc II}}$} & 
\mbox{$l^{{\sc II}}$} & \mbox{$b^{{\sc II}}$} & 
\mbox{$l^{{\sc II}}$} & \mbox{$b^{{\sc II}}$} \\ 
\hline 
 191&   -57.5 &   325 &      -24 &      244 &       21 
& 0.002  & 0.001
& 1.7 & -0.25 & -0.02 \\
\hline
\end{array}$$
\end{table*}
}  

\newcommand\falpha{
\begin{figure}
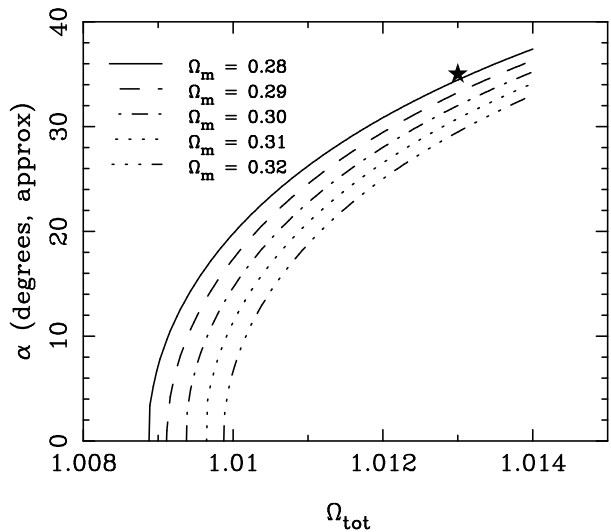

\centering 
\nice{ \centreline{\epsfxsize=8cm
\zzz{\epsfbox[46 28 459 388]{"`gunzip -c alpha5.ps.gz"}}
{\epsfbox[46 28 459 388]{"alpha5.ps"}}  } }
\caption[]{ \mycaptionfont
Dependence of the angular size $\alpha$ of matched circle radii on 
the total density parameter
$\Omtot$ and the non-relativistic matter density 
parameter $\Omm$ for the Poincar\'e dodecahedral hypothesis.
Clearly only slight changes in $\Omtot$ are needed for the circle
radius to change significantly. The values $\Omtot=1.013, \alpha=35\ddeg$
expected by \protect\nocite{LumNat03}{Luminet} {et~al.} (2003) are shown by a star. 
The word `approx' is used in the label to remind the reader that this
is not an exact calculation, since Euclidean geometry is assumed.
}
\label{f-alpha}
\end{figure} 
} 

\newcommand\fomtot{
\begin{figure}
\centering 
\nice{ \centreline{\epsfxsize=8cm
\zzz{\epsfbox[46 28 459 396]{"`gunzip -c omtot5.ps.gz"}}
{\epsfbox[46 28 459 396]{"omtot5.ps"}}  } }
\caption[]{ \mycaptionfont
An inversion of Fig.~\protect\ref{f-alpha}, focussing on the angles
found to be of interest in the WMAP ILC map. 
Independently of $\Omm$, the 
slope in the range $5\ddeg < \alpha < 15\ddeg$ is 
$\mbox{d} \Omtot /\mbox{d}\alpha  \approx 6\e{-5}~\mbox{deg}^{-1}$.
}
\label{f-omtot}
\end{figure} 
} 

\newcommand\fomm{
\begin{figure}
\centering 
\nice{ \centreline{\epsfxsize=8cm
\zzz{\epsfbox[46 28 459 396]{"`gunzip -c omm200.ps.gz"}}
{\epsfbox[46 28 459 396]{"omm200.ps"}}  } }
\caption[]{ \mycaptionfont
Dependence of $\Omm$ on $\alpha$, similarly to 
Fig.~\protect\ref{f-omtot}.
The slope in the range $5\ddeg < \alpha < 15\ddeg$ is 
$\mbox{d} \Omm /\mbox{d}\alpha  \approx 2.2\e{-3}~\mbox{deg}^{-1}$,
with only weak dependence on $\Omtot$ over this interval.
}
\label{f-omm}
\end{figure} 
} 

\newcommand\fcorrm{
\begin{figure}
\centering 
\nice{ \centreline{\epsfxsize=8cm
\zzz{\epsfbox[14 14 430 364]{"`gunzip -c corrm1.ps.gz"}}
{\epsfbox[14 14 430 364]{"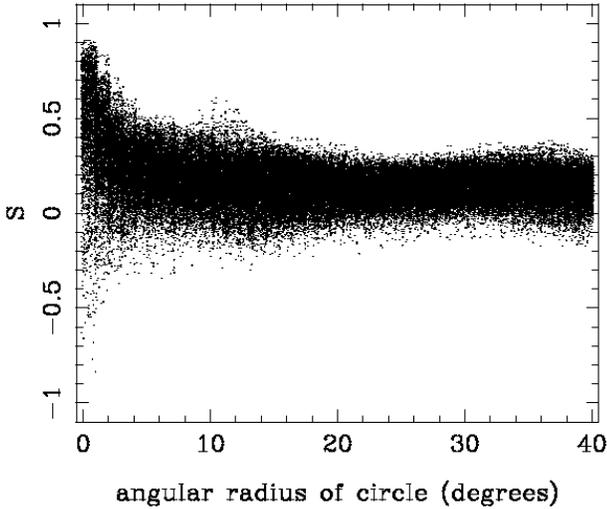"}}  } }
\caption[]{ \mycaptionfont
Values of the correlation statistic $S$
[eq.~(\protect\ref{e-corr})] as a function of matched circle angular
radius $\alpha$ for each pair of six matched circles implied by an
orientation of the fundamental dodecahedron, where the rotation of 
the Clifford translation (screw motion) mapping from a circle to its
image is $-36\ddeg$ (left-handed). 
The values of $\alpha$
are discrete, but a small random offset in $\alpha$ filling the
$0.5\ddeg$ interval between discrete values is used for to better display
individual points. The intervals in $\lII, \bII$ and $\theta$ are $2\ddeg$.
Points with excess correlation appear at $\sim 10\ddeg$.
}
\label{f-corrm}
\end{figure} 
} 

\newcommand\fsigmam{
\begin{figure}
\centering 
\nice{ \centreline{\epsfxsize=8cm
\zzz{\epsfbox[14 14 430 364]{"`gunzip -c sigmam1.ps.gz"}}
{\epsfbox[14 14 430 364]{"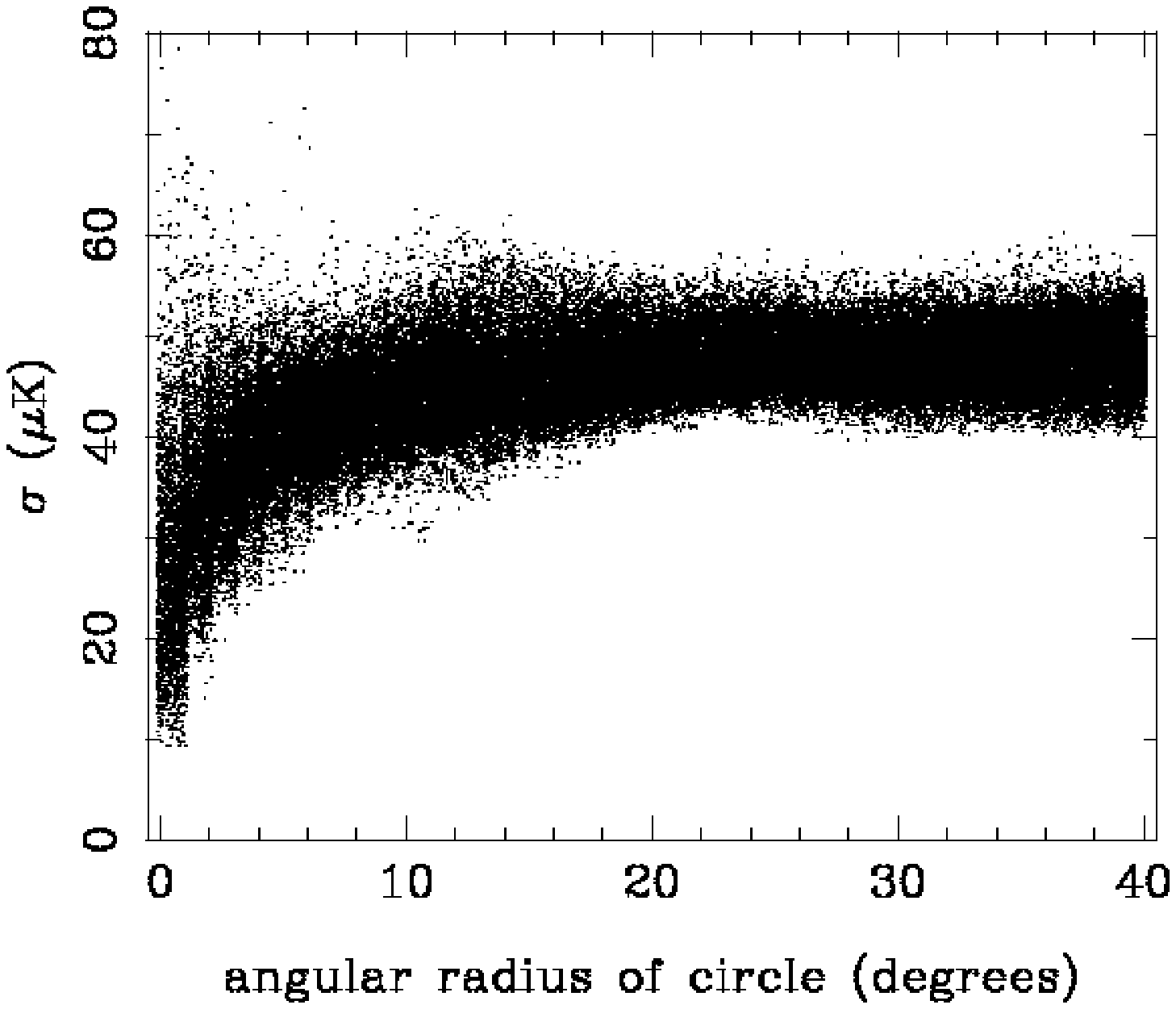"}}  } }
\caption[]{ \mycaptionfont
As for Fig.~\protect\ref{f-corrm}, showing 
values of the difference statistic $\sigma$
[eq.~(\protect\ref{e-sigma})] as a function of matched circle angular
radius $\alpha$ for each pair of six matched circles implied by an
orientation of the fundamental dodecahedron, again for a left-handed
rotation. A cluster of points with smaller differences $\sigma$ than
the general distribution is visible at $\sim 10\ddeg$.
}
\label{f-sigmam}
\end{figure} 
} 

\newcommand\fcorrp{
\begin{figure}
\centering 
\nice{ \centreline{\epsfxsize=8cm
\zzz{\epsfbox[14 14 430 364]{"`gunzip -c corrp1.ps.gz"}}
{\epsfbox[14 14 430 364]{"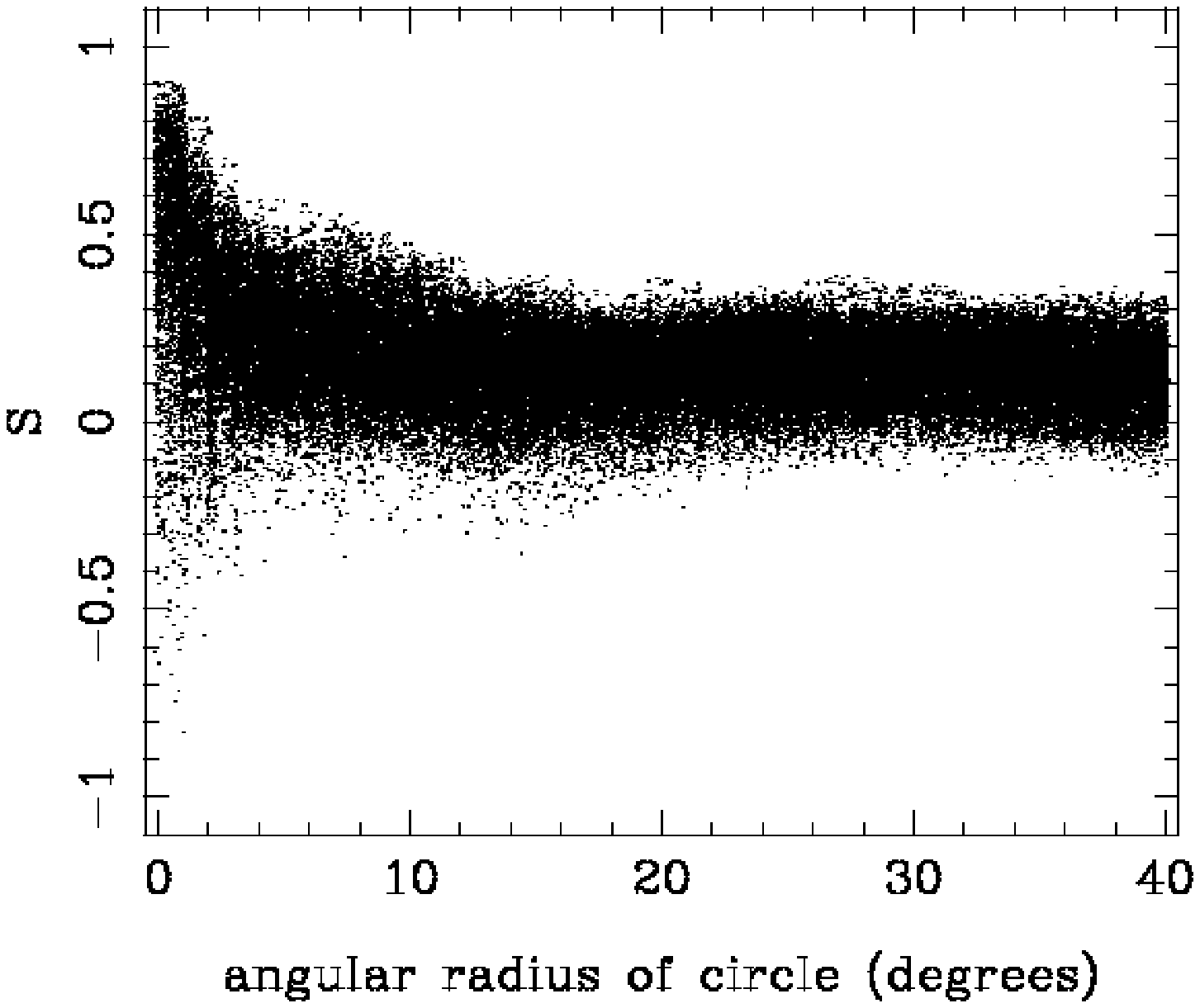"}}  } }
\caption[]{ \mycaptionfont
As for Fig.~\protect\ref{f-corrm}, for a right-handed rotation.
If a signal is present, it should be present for a left-handed
rotation or a right-handed rotation, but not for both.
}
\label{f-corrp}
\end{figure} 
} 

\newcommand\fsigmap{
\begin{figure}
\centering 
\nice{ \centreline{\epsfxsize=8cm
\zzz{\epsfbox[14 14 430 364]{"`gunzip -c sigmap1.ps.gz"}}
{\epsfbox[14 14 430 364]{"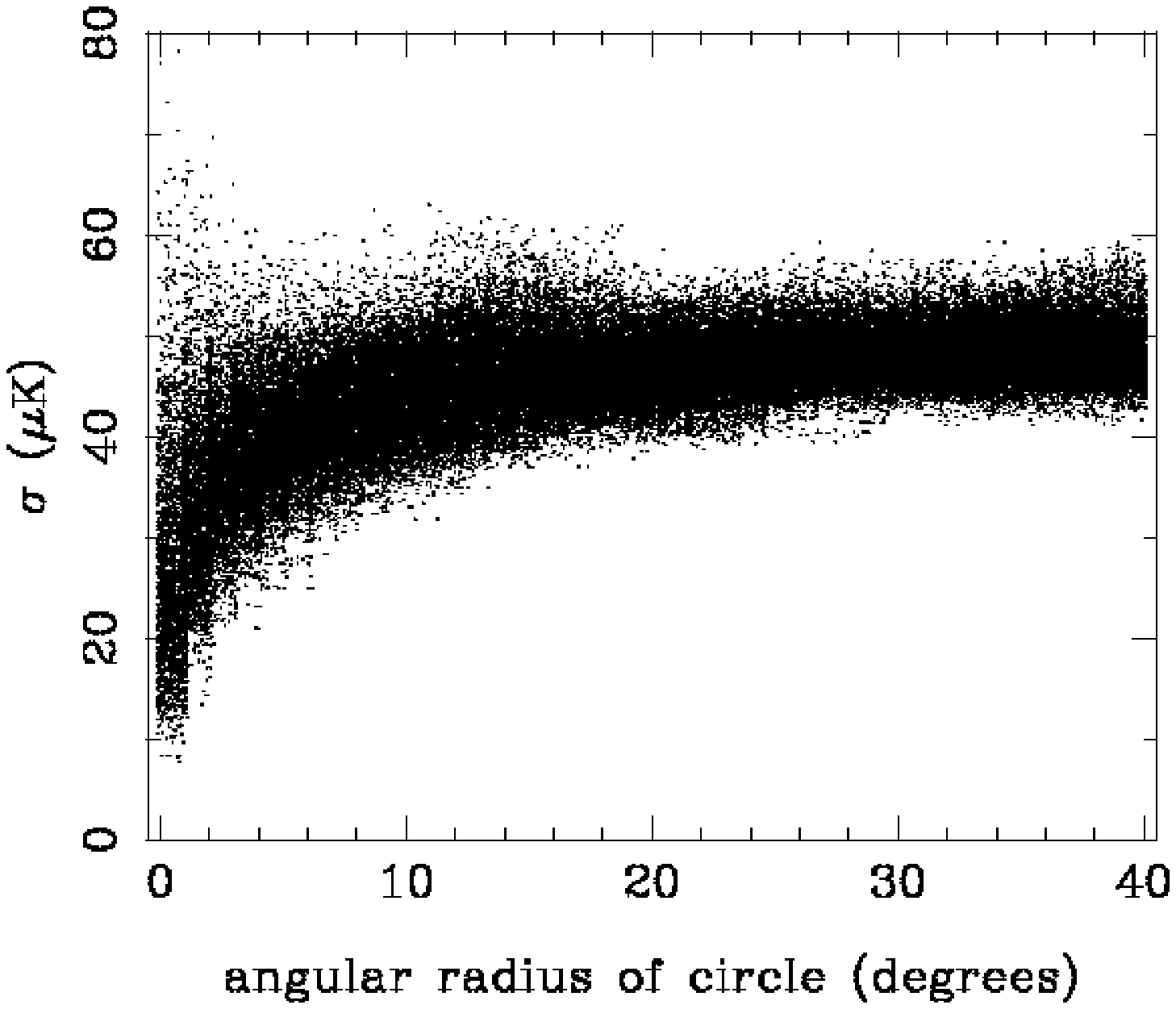"}}  } }
\caption[]{ \mycaptionfont
As for Fig.~\protect\ref{f-sigmam}, for a right-handed rotation.
}
\label{f-sigmap}
\end{figure} 
} 

\newcommand\fcorrz{
\begin{figure}
\centering 
\nice{ \centreline{\epsfxsize=8cm
\zzz{\epsfbox[14 14 430 364]{"`gunzip -c corrz1.ps.gz"}}
{\epsfbox[14 14 430 364]{"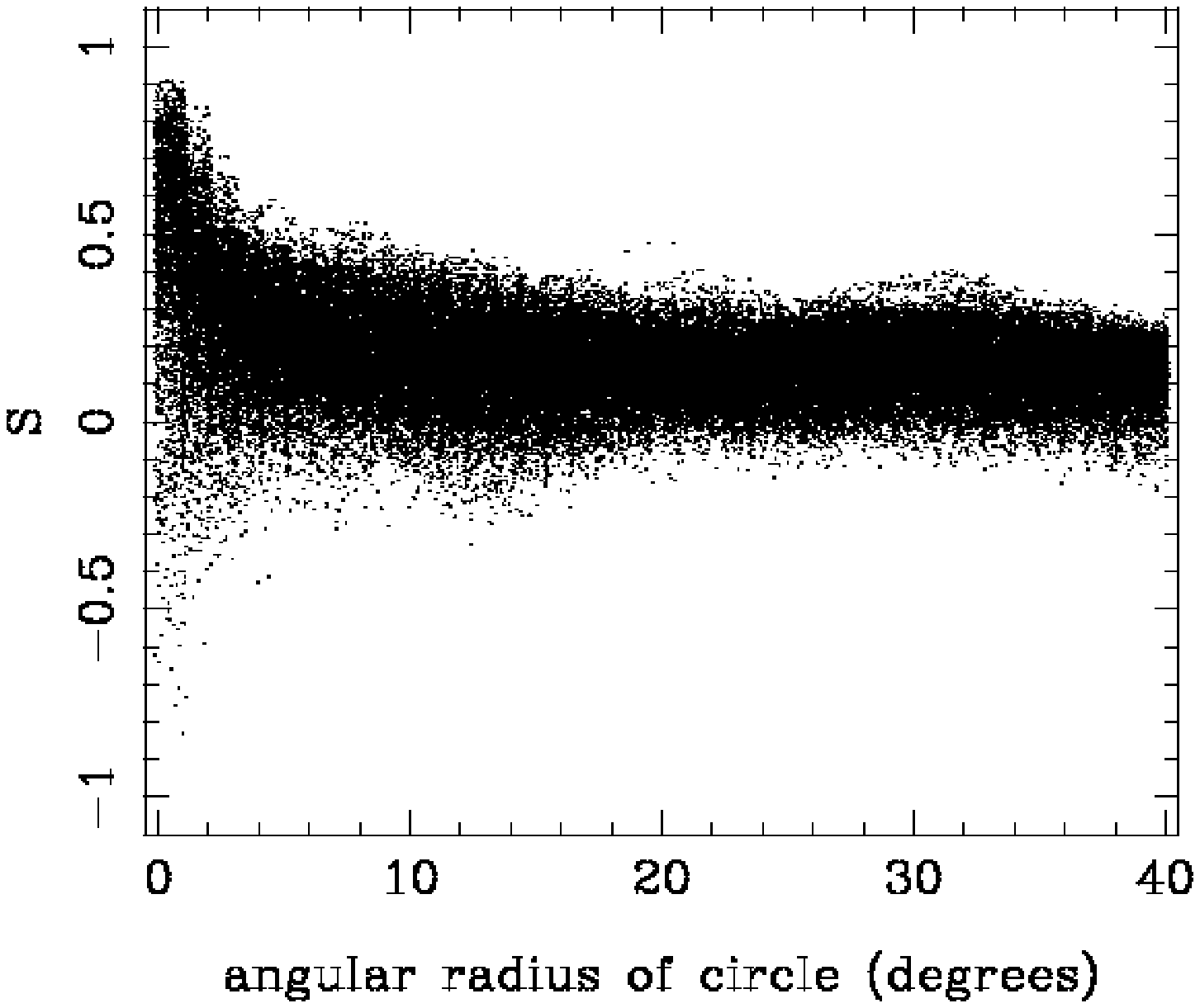"}}  } }
\caption[]{ \mycaptionfont
As for Fig.~\protect\ref{f-corrm}, for a zero rotation. This is
unphysical, so only noise signals should be present.
}
\label{f-corrz}
\end{figure} 
} 

\newcommand\fsigmaz{
\begin{figure}
\centering 
\nice{ \centreline{\epsfxsize=8cm
\zzz{\epsfbox[14 14 430 364]{"`gunzip -c sigmaz1.ps.gz"}}
{\epsfbox[14 14 430 364]{"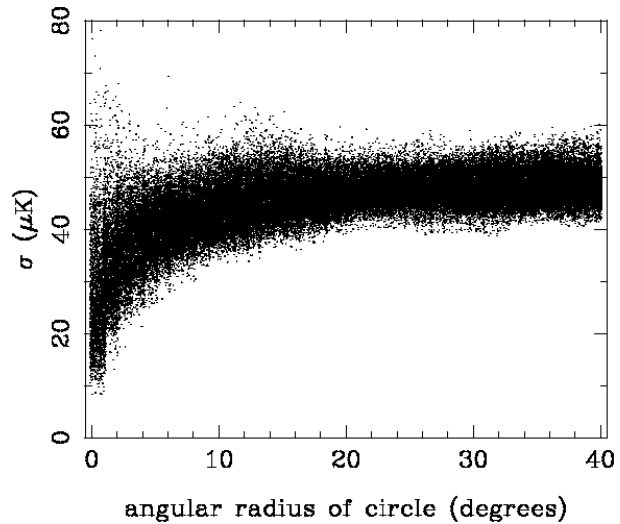"}}  } }
\caption[]{ \mycaptionfont
As for Fig.~\protect\ref{f-sigmam}, for a zero rotation. 
}
\label{f-sigmaz}
\end{figure} 
} 

\newcommand\fcmodem{
\begin{figure}
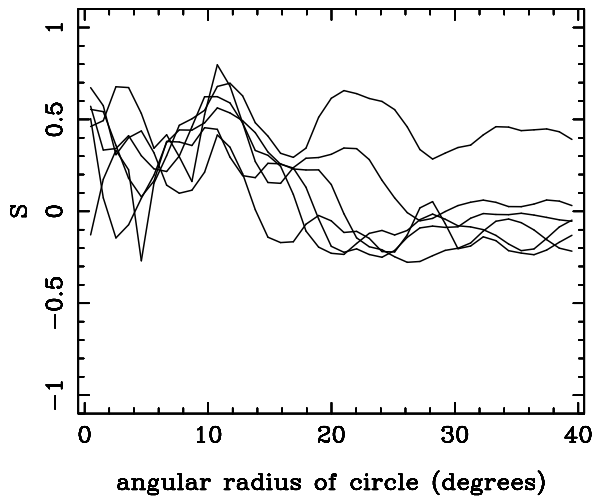

\centering 
\nice{ \centreline{\epsfxsize=8cm
\zzz{\epsfbox[42 34 470 382]{"`gunzip -c c_modem.ps.gz"}}
{\epsfbox[42 34 470 382]{"c_modem.ps"}}  } }
\caption[]{ \mycaptionfont
Values of the correlation statistic $S$ calculated for the 
six individual pairs of circles for dodecahedrons with orientations
in the apparently favoured range [eq.(\ref{e-finedomain})], shown 
as the mode at each circle radius $\alpha$. This is for a left-handed
screw motion. All six pairs have high 
correlations at around $\sim10\ddeg$. 
}
\label{f-c_modem}
\end{figure} 
} 

\newcommand\fcmodep{
\begin{figure}
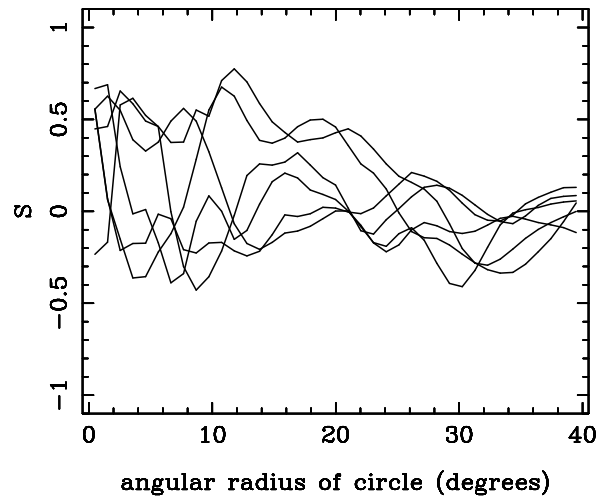

\centering 
\nice{ \centreline{\epsfxsize=8cm
\zzz{\epsfbox[42 34 470 382]{"`gunzip -c c_modep.ps.gz"}}
{\epsfbox[42 34 470 382]{"c_modep.ps"}}  } }
\caption[]{ \mycaptionfont
As for Fig.~\protect\ref{f-c_modem}, for a right-handed
screw motion. Although correlations of some circle pairs are high,
there is no case where all six are high.
}
\label{f-c_modep}
\end{figure} 
} 

\newcommand\fcmodez{
\begin{figure}
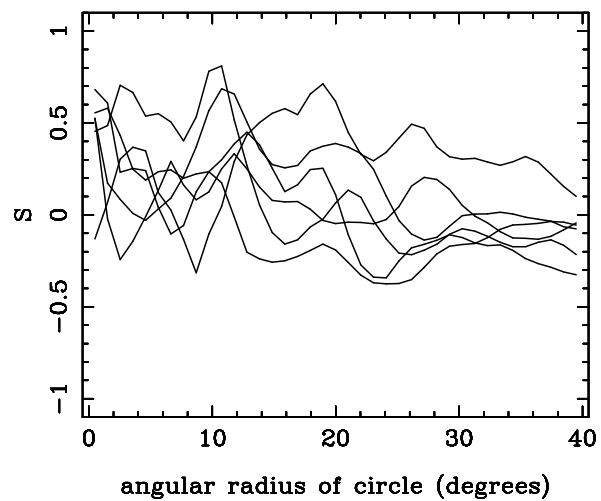

\centering 
\nice{ \centreline{\epsfxsize=8cm
\zzz{\epsfbox[42 34 470 382]{"`gunzip -c c_modez.ps.gz"}}
{\epsfbox[42 34 470 382]{"c_modez.ps"}}  } }
\caption[]{ \mycaptionfont
As for Fig.~\protect\ref{f-c_modem}, for a zero (unphysical)
screw motion. Although correlations of some circle pairs are high,
there is no case where all six are high.
}
\label{f-c_modez}
\end{figure} 
} 

\newcommand\fprobISW{
\begin{figure}
\centering 
\nice{ \centreline{\epsfxsize=8cm
\zzz{\epsfbox[42 34 470 382]{"`gunzip -c roukfig1.eps.gz"}}
{\epsfbox[42 34 470 382]{"roukfig1.eps"}}  } }
\caption[]{ \mycaptionfont
Probability, $P_{\mbox{\rm subs}}$ 
Table~\protect\ref{t-bestshort} as a null hypothesis, that the
{\bf this is just a template} where $1-P=92\%$.
}
\label{f-probISW}
\end{figure} 
} 

\newcommand\fcircleA{ 
\begin{figure*}
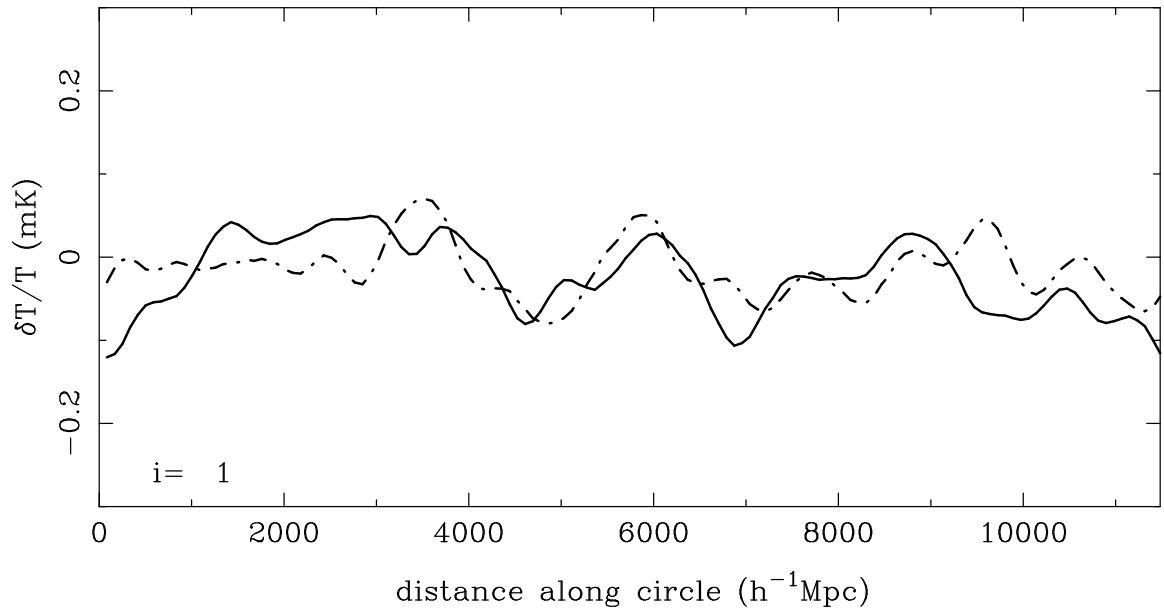

\centering 
\nice{ \centreline{\epsfxsize=8cm
\zzz{\epsfbox[186 10 533 404]{"`gunzip -c circ_1.ps.gz"}}
{\epsfbox[186 10 533 404]{"circ_1.ps"}}  } }
\caption[]{ \mycaptionfont
Temperature fluctuations in the WMAP ILC data
around an $11\ddeg$ radius 
circle centred at position $i=1$ in Table~\protect\ref{t-dodec} 
(continuous curve) and 
its opposite circle (dot-dashed circle), 
matched by a screw motion with a left-handed
rotation of $36\ddeg$ shown against the distance
around each circle for $(\Omm=0.3,\Omega_\Lambda=0.7)$. 
These distances are a slight underestimate of the true distances for
the hypothesis, since Euclidean calculation is used for a `slightly'
positively curved space.
The two curves shown are $\delta T/T$ 
In the general analysis in this paper and 
in this and following plots, point pairs where either point 
lies at a galactic latitudes with 
$|b^{\mbox{{\sc II}}}| < 2\ddeg$ and or within $20\ddeg$ of 
the galactic centre are excluded.
}
\label{f-circleA}
\end{figure*} 
} 

\newcommand\fcircleB{ 
\begin{figure*}
\centering 
\nice{ \centreline{\epsfxsize=8cm
\zzz{\epsfbox[186 10 533 404]{"`gunzip -c circ_2.ps.gz"}}
{\epsfbox[186 10 533 404]{"circ_2.ps"}}  } }
\caption[]{ \mycaptionfont
As for Fig.~\protect\ref{f-circleA}, for 
position $i=2$ in Table~\protect\ref{t-dodec}. 
This is the only one of the six circle pairs for which one of 
the \protect\nocite{WMAPforegrounds}{Bennett} {et~al.} (2003) point sources lies 
within $0.5\ddeg$. This source is GB6~J1635+3808 at $(\lII=61.1\ddeg,
\bII=42.3\ddeg)$ and is shown by a pair of overlapping 
circles at (arbitrarily)
$\delta T/T = 0$. Its sky position is within $0.1\ddeg$ of the
circle shown by the solid curve, suggesting that it is a contaminant
present in the solid curve (\SSS~\ref{s-pointsources}). 
}
\label{f-circleB}
\end{figure*}
} 

\newcommand\fcircleC{ 
\begin{figure*}
\centering 
\nice{ \centreline{\epsfxsize=8cm
\zzz{\epsfbox[186 10 533 404]{"`gunzip -c circ_3.ps.gz"}}
{\epsfbox[186 10 533 404]{"circ_3.ps"}}  } }
\caption[]{ \mycaptionfont
As for Fig.~\protect\ref{f-circleA}, for 
position $i=2$ in Table~\protect\ref{t-dodec}.}
\label{f-circleC}
\end{figure*}
} 

\newcommand\fcircleD{ 
\begin{figure*}
\centering 
\nice{ \centreline{\epsfxsize=8cm
\zzz{\epsfbox[186 10 533 404]{"`gunzip -c circ_4.ps.gz"}}
{\epsfbox[186 10 533 404]{"circ_4.ps"}}  } }
\caption[]{ \mycaptionfont
As for Fig.~\protect\ref{f-circleA}, for 
position $i=2$ in Table~\protect\ref{t-dodec}. The exclusion due to 
the Galaxy mentioned in Fig.~\protect\ref{f-circleA} excludes many
points in this plot.}
\label{f-circleD}
\end{figure*}
} 

\newcommand\fcircleE{ 
\begin{figure*}
\centering 
\nice{ \centreline{\epsfxsize=8cm
\zzz{\epsfbox[186 10 533 404]{"`gunzip -c circ_5.ps.gz"}}
{\epsfbox[186 10 533 404]{"circ_5.ps"}}  } }
\caption[]{ \mycaptionfont
As for Fig.~\protect\ref{f-circleA}, for 
position $i=2$ in Table~\protect\ref{t-dodec}.}
\label{f-circleE}
\end{figure*}
} 

\newcommand\fcircleF{ 
\begin{figure*}
\centering 
\nice{ \centreline{\epsfxsize=8cm
\zzz{\epsfbox[186 10 533 404]{"`gunzip -c circ_6.ps.gz"}}
{\epsfbox[186 10 533 404]{"circ_6.ps"}}  } }
\caption[]{ \mycaptionfont
As for Fig.~\protect\ref{f-circleA}, for 
position $i=2$ in Table~\protect\ref{t-dodec}.}
\label{f-circleF}
\end{figure*}
} 


\section{Introduction}

In the past twelve months, several authors have analysed the possibility that 
the primordial temperature fluctuations in the cosmic microwave
background, as measured in the first-year data of WMAP (Wilkinson
Microwave Anisotropy Probe) satellite 
\nocite{WMAPSpergel}({Spergel} {et~al.} 2003, and accompanying papers), could be better matched by a
perturbed 
Friedmann-Lema\^{\i}tre-Robertson-Walker (FLRW) model in which the global
geometry is multiply connected rather than simply connected.

The possibility of the Universe being multiply connected is first
known to have been suggested by \nocite{Schw00,Schw98}{Schwarzschild} (1900, 1998). 
For recent reviews on cosmic topology, see 
\nocite{LaLu95}{Lachi\`eze-Rey} \& {Luminet} (1995), \nocite{Lum98}{Luminet} (1998), \nocite{Stark98}{Starkman} (1998) and \nocite{LR99}{Luminet} \& {Roukema} (1999). 
For workshop proceedings 
on the subject, see \nocite{Stark98}{Starkman} (1998) and following articles,
and \nocite{BR99}{Blanl{\oe}il} \& {Roukema} (2000). Detection strategies include both two-dimensional
methods (based on temperature fluctuations in the surface of last
scattering) and three-dimensional methods (based on distributions
of gravitationally collapsed objects distributed in three-dimensional
comoving space). For a list and discussion of both two-dimensional and 
three-dimensional methods, 
see Table~2 of \nocite{LR99}{Luminet} \& {Roukema} (1999) and the accompanying discussion. 

The WMAP data has motivated many two-dimensional analyses.

Some of the authors mentioning 
either multiply connected models consistent with 
the WMAP data, or indirect hints of multiple connectedness, include
\nocite{WMAPSpergel,WMAPTegmarkFor,WMAPChiang,WMAPmultipol}{Spergel} {et~al.} (2003); {Tegmark}, {de Oliveira-Costa}, \&  {Hamilton} (2003); {Chiang} {et~al.} (2003); {Copi}, {Huterer}, \& {Starkman} (2003), 
 --- finding low values of low $l$ multipoles or applying a 
multipole vector
analysis. 

However, other authors 
\nocite{FengZhang03,Cline03,Contaldi03,EfstSph03}(e.g.~ {Feng} \& {Zhang} 2003; {Cline}, {Crotty}, \& {Lesgourgues} 2003; {Contaldi} {et~al.} 2003; {Efstathiou} 2003a) 
have suggested various non-topological explanations for the 
low multipole WMAP $C_l$ spectrum, such as double inflation or
other phenomena from early universe physics, or positive curvature.

Yet others \nocite{Naselsky03a,Naselsky03b}({Naselsky}, {Doroshkevich}, \&  {Verkhodanov} 2003, 2004) point out that at least 
the ``internal linear combination'' (ILC) map of the WMAP data contains
non-Poissonian signal, due to foreground residues on large scales. 
On smaller scales, \nocite{Giommi03}{Giommi} \& {Colafrancesco} (2003) find that somewhere between
20-100\% of the signal at spherical harmonic
$l$ values in the range $500 < l < 800$ may be due to blazars.

Attempts to {\em exclude} classes of global geometry models using the WMAP
data include the calculations of \nocite{WMAPTegmarkAgainst}{de Oliveira-Costa} {et~al.} (2004) and of
\nocite{CSSK03}{Cornish} {et~al.} (2004). The latter 
performed a general back-to-back
matched circles analysis 
for angular radii in the range $25-90\ddeg$, using a correlation statistic
for signal detection. They failed to find matched circles 
for a wide class of models, including the torus models, up to a scale
of 16.8{\hGpc}, which \nocite{CSSK03}{Cornish} {et~al.} 2004 state as 24~Gpc, since they
adopt $H_0= 70$\kms Mpc$^{-1}$\footnote{The 
Hubble constant is parametrised 
as $h\equiv H_0/100$km~s$^{-1}$~Mpc$^{-1}.$}.

Disagreement also exists on whether or not the low quadrupole is
really significant in rejecting the infinite flat `concordance' model 
\nocite{EfstNoProb03a,EfstNoProb03b}({Efstathiou} 2003b, 2004).

Possibly one of the strongest claims in favour of a possible detection 
is that by \nocite{LumNat03}{Luminet} {et~al.} (2003), who point out that given
standard assumptions on the statistics of the fluctuations, 
the Poincar\'e dodecahedral model implies a quadrupole and an octupole
very close to those calculated from the WMAP data, for the same
value of the total density parameter, $\Omtot \approx 1.013\pm 0.02$.

In principle, this is excluded by the \nocite{CSSK03}{Cornish} {et~al.} (2004) analysis.

The Poincar\'e dodecahedral model requires positive (spherical)
curvature. \nocite{LumNat03}{Luminet} {et~al.} (2003) favour a total density parameter of
$\Omtot \approx 1.013\pm 0.002$ based on the spherical harmonic statistical
analyses of the WMAP data, with 
non-relativistic matter density parameter, $\Omm=0.28$ 
and cosmological constant $\Omega_\Lambda = 
\Omtot - \Omm$. This implies that points on the surface of last scattering 
which are multiple topological images of single physical points in space-time
should correspond to matched circles \nocite{Corn96,Corn98b}({Cornish}, {Spergel}, \& {Starkman} 1996; {Cornish}, {Spergel}, \&  {Starkman} 1998) which subtend
angular radii at the observer of `about $35\ddeg$' \nocite{LumNat03}({Luminet} {et~al.} 2003). 

Since the angular relations between face centres are identical for
Euclidean and spherical dodecahedra, Euclidean calculations of the
relative positions of circles are sufficient for testing the Poincar\'e
dodecahedral hypothesis. For example, the Euclidean half-angle of
about $31.7\ddeg$ is valid for angular separations of adjacent face
centres in the spherical dodecahedron. Since the adjacent circles
expected by \nocite{LumNat03}{Luminet} {et~al.} (2003) have larger angular radii (subtended at
the observer), this implies that they intersect with the face edges.

However, as is shown in \SSS\ref{s-notthirtyfive}, this angular 
radius is extremely sensitive to the value of $\Omtot$. For example,
keeping $\Omm=0.28$ fixed, it is sufficient to decrease $\Omtot$ to 
$\Omtot=1.009$ to bring the angular radius to nearly zero. 

This implies that a Poincar\'e dodecahedral signal may have been
missed by \nocite{CSSK03}{Cornish} {et~al.} (2004) because they did not explore the part of 
parameter space for small matched circles.

In this paper, the missing part of parameter space is investigated.  A
matched circles analysis specifically designed to detect dodecahedral
patterns of matched circles is performed over angular radii in the
range $1-40\ddeg$ on 
Internal Linear
Combination map (ILC) of 
the one-year WMAP data. The WMAP data 
are briefly discussed in \SSS\ref{s-wmap}. 
While the ILC is unlikely to be ideal for the studies of $C_l$ 
statistics, and some authors (cited above) claim correlations with
foregrounds, it is hard to see how any signal mimicking matched circles
oriented in a dodecahedral pattern could be imposed, either by 
the construction method of the ILC or by foregrounds.

Signal detection is
attempted via a correlation statistic and
an rms difference statistic, similarly to \nocite{Rouk00a,Rouk00b}{Roukema} (2000b, 2000a).
Extreme value distributions of these statistics between
a right-handed rotation when matching circles, 
a left-handed rotation, and a zero rotation. A genuine signal should appear
for either the right-handed or left-handed rotation, but not both, and
should not appear for the zero rotation. 

The relation between circle angular radius ($\alpha$) 
and local cosmological 
parameters ($\Omm, \Omtot$) and the statistics used are presented in 
\SSS\ref{s-method}. 

The tentative detection of dodecahedrally distributed matched circles
with $\alpha \approx 10\ddeg$ and analysis of their statistical significance
are presented in
\SSS\ref{s-results}. Further discussions and conclusions are made in
\SSS\ref{s-conclu}.

For reviews on cosmological topology, see 
\nocite{LaLu95}{Lachi\`eze-Rey} \& {Luminet} (1995), \nocite{Lum98}{Luminet} (1998), \nocite{Stark98}{Starkman} (1998) and \nocite{LR99}{Luminet} \& {Roukema} (1999). 
For workshop proceedings 
on the subject, see \nocite{Stark98}{Starkman} (1998) and the following articles,
and \nocite{BR99}{Blanl{\oe}il} \& {Roukema} (2000).
For a list and discussion of both two-dimensional and 
three-dimensional methods, 
see Table~2 of \nocite{LR99}{Luminet} \& {Roukema} (1999) and the accompanying discussion. 
The reader should be reminded that while microwave background data 
is still the most popular for topology analyses, 
considerable work in three-dimensional methods has been carried 
out, including, e.g., 
\nocite{LLL96,Rouk96,FagG97,RE97,RB98,Gomero99a,LLU98,FagG99a,ULL99a,FagG99b,Gomero99b,Gomero99c}{Lehoucq}, {Lachi\`eze-Rey}, \&  {Luminet} (1996); {Roukema} (1996); {Fagundes} \& {Gausmann} (1998b); {Roukema} \& {Edge} (1997); {Roukema} \& {Blanl{\oe}il} (1998); {Gomero}, {Reboucas}, \&  {Teixeira} (2002); {Lehoucq}, {Luminet}, \& {Uzan} (1999); {Fagundes} \& {Gausmann} (1998a); {Uzan}, {Lehoucq}, \& {Luminet} (1999); {Fagundes} \& {Gausmann} (1999); {Gomero}, {Reboucas}, \&  {Teixeira} (2000, 2001).

For background on spherical multiply connected spaces, apart from the 
recent analysis by 
\nocite{LumNat03}{Luminet} {et~al.} (2003), see \nocite{GausSph01,LehSph02,RiazSph03}{Gausmann} {et~al.} (2001); {Lehoucq} {et~al.} (2002); {Riazuelo} {et~al.} (2003) for extremely
thorough, in-depth
mathematical background directly related to the cosmological context.
\postrefereechanges{
For a general background on geometry and topology, see e.g. 
\nocite{Weeks2001}{Weeks} (2001).
}

\falpha

\fomtot

\fomm

Comoving coordinates are
used when discussing distances 
(i.e. `proper distances', \nocite{Wein72}{Weinberg} 1972, equivalent to `conformal time'
if $c=1$). 

\section{Observations} \label{s-wmap}

See \nocite{WMAPSpergel}{Spergel} {et~al.} (2003) and accompanying papers for an introduction 
to the WMAP data.

While some authors make their own combinations of the WMAP maps in
individual frequencies, it is hard to see how any signal mimicking
matched circles oriented in a dodecahedral pattern could be imposed,
by the method according to which the Internal Linear Combination 
(ILC) map is constructed. So this is the data set used here.

However, in order to obtain conservative results, all points
within $2\ddeg$ of the galactic plane or $20\ddeg$ of the galactic 
centre are removed from the analysis.

Moreover, a smoothing of $1\ddeg$ HWHM is applied for the main 
calculation, since if a signal
is present, it should be present both on super-degree scales and on
sub-degree scales, but should be easier to detect on the larger scales,
because the search calculations can be made much faster, because the 
na\"{\i}ve Sachs-Wolfe dominates over the Doppler effect on larger scales,
and because there is likely to be less influence from subtle effects
which need to be taken into account as correction factors, e.g. the
special relativistic correction for the observer's motion with respect
to the microwave background, \nocite{LevinTwins01,UzanTwins02}{Barrow} \& {Levin} (2001); {Uzan} {et~al.} (2002).

\section{Method} \label{s-method}

\subsection{The Identified Circles Principle} \label{s-matchedc}

The identified circles principle was first published by 
Cornish, Spergel \& Starkman
(\nocite{Corn96,Corn98b} 1996, 1998).
This defines the set of multiply topologically imaged points for
a given manifold. 

This set can be generated 
by considering copies of the observer in the covering space 
placed at distances
less than the horizon diameter from the observer. The 
intersection of the two surfaces of last scattering (spheres) of 
the observer and a copy of the observer is a circle. Since
the copy of the observer is physically identical to the observer,
what appears to be two observers looking at
one circle  
is equivalent to one observer looking at two circles.

Hence, each pair of virtual copies of the observer in the covering
space (separated by less than the diameter of the surface of last
scattering) implies a pair of circles on the surface of last scattering
which correspond to identical points in space-time.

\subsection{Relating circle angular radius $\alpha$ to $\Omm$, $\Omtot$} 
\label{s-notthirtyfive}

As mentioned above, Euclidean geometry is sufficient for calculating
the positions of matched circles for the Poincar\'e dodecahedral 
hypothesis of \nocite{LumNat03}{Luminet} {et~al.} (2003), even though the model is for positive
curvature. 
%
\postrefereechanges{
Errors would occur only if, for example, the 
angular radius of a circle, $\alpha$,
is converted 
into a comoving distance in megaparsecs, or if the ratio of 
the inverse cosine of the ratio of the in-radius, $r_{-}$, to
the distance to the surface of last scattering, $\rSLS$, is used 
to infer the
angular radius of a matched circle, i.e. 
\begin{equation}
\alpha \approx \cos^{-1} \left( {r_{-} \over \rSLS} \right) 
\label{e-alphaestimate}
\end{equation}
}

Since the values of $\Omtot$ of interest are very close to unity,
either of these calculations should only lead to small errors. 

In fact, a much bigger uncertainty is due to the sensitivity of $\alpha$
to the values of $\Omtot$ and $\Omm$.

The curvature radius can be defined in a strict sense, without taking an
absolute value of the curvature, as
\begin{equation}
R_C = { c \over H_0} { 1 \over \sqrt{ \Omtot - 1 } }.
\label{e-rcdefn}
\end{equation}
This has the advantage of showing that the transition from `slightly' 
hyperbolic space to flat space to spherical space, i.e. from 
$\Omtot < 1$ to $\Omtot =1 $ to $\Omtot > 1$ starts from imaginary 
values of the curvature radius, increases to larger and larger imaginary
curvature radii without bounds, goes through a singularity, and decreases
from unbounded large, real values to smaller and smaller real values 
as $\Omtot$ increases through positive values.

It is therefore unsurprising that other physical quantities related to
the curvature radius may show rapid changes near the flatness limit.

To illustrate the relation between $\alpha,$ $\Omtot$ and $\Omm$, 
the value of $r_{-}$, the in-radius of the 
fundamental domain of the Poincar\'e dodecahedron, expressed as 
\begin{equation}
r_{-} = 0.31 R_C
\label{e-rindefn}
\end{equation}
is used, along with eq.~(\ref{e-alphaestimate}) and the comoving 
distance to the surface of last scattering
\begin{equation}
\rSLS = r(\Omm,\Omtot,H_0, z = 1100).
\label{e-rslsdefn}
\end{equation}

Fig.~\ref{f-alpha} shows how a few tenths of a percent drop in $\Omtot$ 
are sufficient to increase the curvature radius so that 
the in-radius of the fundamental dodecahedron becomes as large as the
surface of last scattering, reducing the circle size from $\approx 35\ddeg$
to zero.

So, while \nocite{LumNat03}{Luminet} {et~al.} (2003)'s statistical estimates of $C_l$ values do 
suggest that $\Omtot \approx 1.013$ and $\alpha\approx 35\ddeg$, it 
is clear that smaller $\Omtot$ values and circle sizes should be
investigated.

In Figs~\ref{f-omtot} and \ref{f-omm}, the dependence of $\Omtot$ and
$\Omm$ on $\alpha$ in the angular range found to be of interest in the
WMAP data is illustrated. Given the fact that the Universe {\em is} 
anisotropic (we clearly do {\em not} live in a true FLRW universe; 
we live in a perturbed FLRW universe) some variation in the size of circles
could be expected due to perturbations in the values of $\Omm$ and
$\Omtot$. 

The variation shown in Fig.~\ref{f-omm},
\begin{equation}
\mbox{d} \Omm /\mbox{d}\alpha  \approx 2.2\e{-3}~\mbox{deg}^{-1},
\label{e-dommdalpha}
\end{equation}
is clearly much larger than temperature fluctuations on scales of
about a radian ($l \sim 2-3$). 

However, the variation in $\Omtot$,
\begin{equation}
\mbox{d} \Omtot /\mbox{d}\alpha  \approx 6\e{-5}~\mbox{deg}^{-1},
\label{e-domtotdalpha}
\end{equation}
is only slightly larger than that indicated by temperature fluctuations.
\postrefereechanges{
Moreover, the unknown physics of the cosmological constant (or 
quintessence) makes it hard to {\em a priori} exclude any fluctuations
in $\Omega_\Lambda$, especially with an amplitude so close to that of 
matter fluctuations.
}

Hence, a variation in circle size of about one degree would be consistent
with fluctuations in $\Omega_\Lambda$ on scales of about a radian of
amplitude about $6\e{-5}$.

Along with restrictions in practical computing power, this motivates
a search strategy and smoothing on a scale of about one degree.

\subsection{Parameter space} \label{s-parspace}

The space of possible orientations of a dodecahedron (centred on the
observer) and possible circle matches is defined by five parameters: 
three for the orientation of the dodecahedron, one for the 
mapping between faces, and one for the circle radius.

The orientation of the dodecahedron depends on 
three continuous free parameters. The definition used
here includes the
galactic longitude and latitude $(\lII,\bII)$ of one face centre, and
a third parameter allowing a rotation of up to $\theta = 2\pi/5 =
72\ddeg$ around the axis defined by the first centre and its opposite.
The zero point of the rotation $\theta$ is arbitrary.

A given triple $(\lII,\bII,\theta)$ can equivalently be written 
as a quadruple $(\lII_1,\bII_1,\lII_2,\bII_2)$ representing two 
non-opposite face centres, which uniquely define 
the same orientation.

The mapping of one copy of the fundamental domain to
the next, or what is most interesting in the present case, 
the matching of one face of the dodecahedron to its matching face,
has one free 
binary parameter, $\gamma,$ the handed-ness of the screw motion of the
mapping (which is a Clifford translation). 
For a given triple $(\lII,\bII,\theta)$, the screw motion
can either be left-handed or right-handed. The amplitude of the
screw motion is $\pi/5 = 36\ddeg$, equal to the in-diameter of
the fundamental domain in units of the curvature radius. 

For a rotation of $\gamma \pi/5,$ the only physically valid values
are $\gamma = \pm 1$.

The remaining parameter is that discussed above, the angular radius
of a circle, $\alpha$.

Putting these together, these can be written as five-dimensional parameter
space: 
\begin{equation}
(\lII,\bII,\theta, \gamma, \alpha),
\label{e-parspace}
\end{equation}
where four parameters are continuous variables and 
one of the parameters, $\gamma,$ has only two 
physically possible values, $\pm1$.

The binary parameter provides an easy method for providing a control
test which should demonstrate the likely distribution of values of 
statistical measures of circle identity for false matches. 

By setting the Clifford translation as a translation plus a zero
rotation, i.e. setting $\gamma=0$, 
rather than a translation plus a $\pm36\ddeg$ rotation, 
an extra 4-plane in parameter space, in which there is sure to be no
genuine signal, is provided. 

So, this extra 4-plane is tested here. 

Strictly speaking, values of $(\lII,\bII)$ only need to be searched
over one-twelth of the sphere, since outside of one spherical pentagon 
of the spherical dodecahedron, redundant points would be searched.
For simplicity, the search is made for $\bII > +50\ddeg$, so that 
some redundant points are expected. 

The full
parameter search space is:
\begin{eqnarray}
0\ddeg \le \lII \le 360\ddeg \nonumber \\
50\ddeg \le \bII \le 90\ddeg \nonumber \\
0\ddeg \le \theta \le 72\ddeg \nonumber \\
\gamma \in \left\{0,-1,+1\right\} \nonumber \\
1\ddeg \le \alpha \le 40\ddeg, 
\label{e-parspacelimits}
\end{eqnarray}
where the zero screw motion is a control experiment which can only 
give false detections.

\subsection{Correlation and difference statistics}

For a given triple $(\lII,\bII,\theta)$, $\alpha$, and the handed-ness
of the screw motion, there are six pairs of circles which should have
identical temperature fluctuations if there were only the na\"{\i}ve
Sachs-Wolfe effect, no foregrounds and no other statistical or systematic
sources of noise.

Here, the statistic of \nocite{Corn98b}{Cornish} {et~al.} (1998) is used as the correlator,
\begin{equation}
S\equiv { \left<
         { 2 \left({\delta T \over T}\right)_i 
                  \left({\delta T \over T}\right)_j }\right> 
     \over { 
\left< \left({\delta T \over T}\right)_i^2 +
    \left({\delta T \over T}\right)_j^2  \right> } },
\label{e-corr}
\end{equation}
where $i$ and $j$ are the hypothetically multiply imaged locations 
in space-time. The mean is calculated using linearly interpolated
temperature fluctuations in the ILC at intervals of $0.5$ 
great circle degrees.

This is essentially 
a two-point autocorrelation function normalised by the variance.

The other statistic used is an rms difference statistic:
\begin{equation}
\sigma^2 \equiv { \left<
         { \left[ \left({\delta T \over T}\right)_i -
                  \left({\delta T \over T}\right)_j   \right]^2 }
                   \right>, }
\label{e-sigma}
\end{equation}
using the same notation.

Both statistics are means for all six circle pairs for the given
point in parameter space $(\lII,\bII,\theta, \pm1, \alpha)$.

For genuine matched circles, $S$ should be {\em larger} than for surrounding
regions in parameter space, while $\sigma$ should be {\em smaller} than
for surrounding regions.

\section{Results} \label{s-results}

\subsection{Search}

The full parameter space [eq.~(\ref{e-parspacelimits})] is searched 
with a resolution of $2\ddeg$. Since the parameter space is large,
not all values of $S$ and $\sigma$ are stored. Instead, over 
the $\theta$ dimension, values of $S$ and $\sigma$ are stored for the 
two values of $\theta$ which maximise $S$ and minimise $\sigma$ 
respectively. 
These two values of $\theta$ may be, but are not
necessarily, identical. \nocite{CSSK03}{Cornish} {et~al.} (2004) write $\Smax$ to indicate that 
$S$ is a maximum value, not a mean value, but this is not strictly 
identical to the value used here, since values of both $S$ and 
$\sigma$ are stored for both the cases of maximum $S$ and minimum $\sigma$.

Taking the same approach as \nocite{CSSK03}{Cornish} {et~al.} (2004), the values of $S$ and 
$\sigma$ are plotted as a function of $\alpha$, for the Clifford
translations of both orientations, i.e. rotations of $\pm36\ddeg$ and
for a zero rotation. 

These values are shown in Figs~\ref{f-corrm}--\ref{f-sigmaz}.

\fcorrm
\fcorrp
\fcorrz

There is clearly a cluster of points at $\sim10\ddeg$ which show an
excessly high values of $S$ (Fig.~\ref{f-corrm}) and excessively low
values of $\sigma$ (Fig.~\ref{f-sigmam}) for the left-handed 
rotation, in contrast to the general distributions of these
statistics in the right-handed and zero (unphysical) cases.

The general increase in the value of $S$ as $\alpha$ decreases
is similar to its behaviour in the figures of \nocite{CSSK03}{Cornish} {et~al.} (2004), and
consistently, there is 
a corresponding decrease in $\sigma$ values as $\alpha$ decreases.

Could this cluster of points indicate a real signal?

If they correspond to a real signal, spread out over about a degree
due to smoothing of the ILC map, to origin of the signal from 
the na\"{\i}ve Sachs-Wolfe effect at super-degree
scales, and possibly due to intrisic variations in $\Omega_\Lambda$ 
of $\Delta \Omega_\Lambda \sim 6\e{-5}$ [eq.~(\ref{e-domtotdalpha})], then 
\begin{list}{(\roman{enumi})}{\usecounter{enumi}}
\item they should represent 
adjacent points in parameter space
$(\lII,\bII,\theta, \pm1, \alpha)$, and 
\item each of the six individual
circle pairs should show excess correlations and low differences 
near the optimal point.
\end{list}

\subsection{(i) location of the $\sim10\ddeg$ excess}

The twenty points with the highest values of $S$ in the range 
$5\ddeg < \alpha < 15\ddeg$ 
all lie in one of the two ranges 
\begin{eqnarray}
\lII &=& 253 \pm 4\ddeg \nonumber \\
\bII &=& 64 \pm 1\ddeg \nonumber \\
\theta &=& 57.0  \ddeg \nonumber \\
&&\mbox{(the point with the lowest $S$ value is at $59.0\ddeg$)} \nonumber \\
\label{e-solutionone}
\end{eqnarray}
or
\begin{eqnarray}
\lII &=& 50 \pm 3.5\ddeg \nonumber \\
\bII &=& 51.0 \ddeg \nonumber \\
\theta &=& 25 \pm 2 \ddeg.
\label{e-solutiontwo}
\end{eqnarray}

These two triples $(\lII,\bII,\theta)$ correspond to the 
same orientation of the dodecahedron --- they are two face centres 
for a single orientation.
As mentioned above, the full parameter space is covered a little more 
than once, in order to be sure that all orientations are tested.

The difference statistic $\sigma$ varies more rapidly with $\alpha$, 
so a narrower range around $\sim10\ddeg$ is required to see if $\sigma$
isolates similar points to $S$.

Of the twenty points with the lowest values of $\sigma$ in the range 
$8\ddeg < \alpha < 15\ddeg$,  
all but four lie in the ranges of (\ref{e-solutionone}) and 
(\ref{e-solutiontwo}), except for four points with
\begin{eqnarray}
\lII &=& 57.5 \ddeg \nonumber \\
\bII &=& 80.0 \pm 1 \ddeg \nonumber \\
\theta &=& 67 \pm 2 \ddeg.
\label{e-badsolution}
\end{eqnarray}
However, the four points of (\ref{e-badsolution}) have correlations
$0.27 \le S \le 0.30$, much lower than for the other sixteen points,
which have $0.50 \le S \le 0.61.$ This solution is not a face centre
of the same dodecahedron as (\ref{e-solutionone}) and 
(\ref{e-solutiontwo}).

\fsigmam
\fsigmap
\fsigmaz

Clearly, only the solution (\ref{e-solutionone}) and 
(\ref{e-solutiontwo}) is simultaneously indicated by both statistics.

\subsection{(ii) individual circle pairs} \label{s-individualpairs}

If the $\sim 10\ddeg$ signal is really due to the Poincar\'e
dodecahedral topology, then it should appear independently in each of
the six circle pairs. If it is due to non-Gaussian and/or non-Poissonian
properties of the ILC, it is unlikely to appear in all six pairs.

\tdodec

To test this, the correlation statistic $S$ is calculated separately
for each of the six individual pairs within the range
\begin{eqnarray}
\lII &=& 253 \pm 5\ddeg \\
\bII &=& 65 \pm 1\ddeg \\
\theta &=& 57.0 \pm 1 \ddeg, 
\label{e-finedomain}
\end{eqnarray}
with a $0.5\ddeg$ resolution.

Since the domain in most parameters is small, the mode for any given
$\alpha$ value is plotted rather than a scatter plot of the individual
values.

Figs~\ref{f-c_modem},\ref{f-c_modep} and \ref{f-c_modez} clearly show
that the peak correlation at around $10\ddeg$ is present in all six
circle pairs for the left-handed screw motion.

Correlations for a few individual peaks reach similar values for the
right-handed and zero (unphysical) screw motions, but not for all six
pairs simultaneously.

The maximum of the individual curves in Fig.~\ref{f-c_modem} is used
to estimate the circle radius. 

As shown in Table~\ref{t-dodec}, these radii all agree to within
one degree, i.e. $\alpha = 11\pm1 \ddeg$.

\subsection{Matched circles}

What are the actual values of the temperature fluctuations along the
identified circles? The identified circles for the central values
listed in eq.~(\ref{e-finedomain}) are shown in
Figs~\ref{f-circleA}--\ref{f-circleF}.

These clearly show that temperature values around these circles
are highly correlated.

\subsection{Point source catalogue} \label{s-pointsources}

If the matched circles are due to topology, then any foreground 
point sources close to them in celestial position should be present in 
one circle but not the other.

Of the 208 point sources listed in the WMAP point source catalogue
\nocite{WMAPforegrounds}({Bennett} {et~al.} 2003), only one is within $0.5\ddeg$ of the 
matched circle solution. 
This source is GB6~J1635+3808 at $(\lII=61.1\ddeg,
\bII=42.3\ddeg)$ and is located within $0.1\ddeg$ of the
circle shown by the solid curve in Fig.~\ref{f-circleB}.

The point source appears to be present in the curve (at position 
along the circle of about 1000{\hMpc}, 
shown by a pair of overlapping circles).
A removal of flux in order to correct for the presence of this point
source would improve the match between the circles.

\section{Discussion and Conclusions} \label{s-conclu}

It is clearly premature to claim a highly significant 
detection of the topology of 
the Universe based on just one simple analysis of the first year WMAP 
ILC cosmic microwave background map.

However, the plots are striking and it seems prudent to release
them to the scientific community while a companion paper is prepared
with formal statistical analyses.

Whether or not the matched circles found are just coincidence or due
to global geometry, it is clear that temperature fluctuations around
12 dodecahedrally spaced circles of radius $11\pm1\ddeg$ in the WMAP
ILC map correlate unusually well in their respective pairs when a
phase shift of $36\ddeg$, corresponding to a left-handed screw motion,
is applied.

\fcmodem
\fcmodep
\fcmodez

If simulations can show that this is a fairly likely occurrence due
to Gaussian fluctuations in an infinite flat universe, then this would
show that genuine matched circles will be even harder to distinguish
from spurious detections than was previously thought.

However, \nocite{Vielva03}{Vielva} {et~al.} (2003) have found non-Gaussian fluctuations at
about $10\ddeg$ - extremely close to radius of 
the matched circle radius. While it is not obvious how this non-Gaussianity
detection should relate to the radius of matched circles, since correlations
are between circles, not along individual circles, this would also 
complicate simulations, since they would need to be consistent with
observational analyses like this one.

\nocite{CSSK03}{Cornish} {et~al.} (2004) avoided circles of radii smaller than $25\ddeg$ 
because of the risk of false positives, and carried out extensive
simulations of what signal would be expected from a genuinely multiply
connected universe.

However, these simulations risk the problem of cosmic variance 
and logical circularity. If the Universe really is detectably multiply
connected, then it would be fairly reasonable that the density perturbations
(eigenmodes) on the largest scales are somewhat affected by the global
physics of the Universe, so that modelling them on the basis of 
Gaussianity and random phases may lead to statistical statements which 
are incorrect, because they talk about an ensemble of likely universes
with different statistical properties to the real Universe. 

\fcircleA
\fcircleB
\fcircleC
\fcircleD
\fcircleE
\fcircleF

The cosmic variance problem is that our actual Universe is just one
realisation --- the number of perturbations on large scales are too small
for the large number theorem to make statistics of ensembles valid.

An apparently strong heuristic argument against these matches being 
physical is the need for extreme fine-tuning. If the claimed matched
circles are due to topology, then the in-radius of the Universe
happens by chance to be about $\cos(11\ddeg) \approx 98\%$ of the
distance to the surface of last scattering. Intuitively, this is 
difficult to accept.

However, as pointed out by \nocite{LumNat03}{Luminet} {et~al.} (2003), in the case of a
positively curved Universe, especially if $\Omtot \approx 1.01$, there
is necessarily a fine-tuning, since the curvature scale is only about
three times the matter-dominated horizon size.

Moreover, the non-zero cosmological constant, $\Omega_\Lambda \approx
0.7$, is well established observationally (though see
\nocite{BlanchSarkar03,VaucBlanch03}{Blanchard} {et~al.} (2003); {Vauclair} {et~al.} (2003) for a minority viewpoint) and 
definitely requires fine-tuning of some sort.

Attempts have been made to link a non-zero cosmological constant with
detectable cosmic topology, but so far no obvious successes have been
found.

Independently of theoretical arguments, the results presented in this
paper are testable by several observationally based
methods which do not require assumptions
on hypothetical statistical ensembles of universes, and each 
should potentially be able to 
improve the signal if it is cosmological in origin:
\begin{list}{(\roman{enumi})}{\usecounter{enumi}}
\item an attempt to separate out the na\"{\i}ve Sachs-Wolfe effect, 
the doppler component, and the integrate Sachs-Wolfe effect, and 
analysis at higher resolution
\item improved removal of foregrounds, either independently of the
hypothesis, or using the matched circles hypothesis to predict
foregrounds which intervene in one circle but not the other 
(cf. \SSS\ref{s-pointsources})
\item polarisation data from Planck.
\end{list}

Similarly to the prediction by \nocite{LumNat03}{Luminet} {et~al.} (2003) 
that `$\Omtot \approx 1.013 > 1$', Fig.~\ref{f-alpha} shows
that for $\Omm = 0.28\pm0.02,$ the total density parameter must be
$\Omtot \approx 1.010 \pm 0.001$ for these matched circles of
radius $11\pm1\ddeg$ to be
cosmological in origin. A larger uncertainty in $\Omm$ would 
correspondingly increase the uncertainty in $\Omtot$, but the 
prediction that $\Omtot > 1$ remains: the fundamental dodecahedron 
of the Poincar\'e dodecahedral manifold is that of a manifold of positive 
curvature.

\section*{Acknowledgments}

B.F.R. is extremely grateful to Roman Juszkiewicz, Marek Demia\'nski,
Bronek Rudak and Andrzej Kus, who enabled him to continue doing
cosmology research.
Use was made of the WMAP data
\url{http://lambda.gsfc.nasa.gov/product/} and of the
Centre de Donn\'ees astronomiques de Strasbourg 
\url{http://cdsads.u-strasbg.fr}.
Some of the results in this paper have been derived using the 
HEALPix 
\nocite{Healpix98}({G\'orski}, {Hivon}, \&  {Wandelt} 1999)\footnote{see 
\protect\url{http://www.eso.org/science/healpix/}}.
SB acknowledges support from KBN Grant 1P03D 012 26.

\subm{ \clearpage }

\nice{

}


\end{document}